\documentclass[prd,preprint,superscriptaddress,amsmath,amssymb,nofootinbib]{revtex4}
\usepackage{graphicx}
\usepackage{dcolumn}
\usepackage{bm}
\usepackage{amssymb}
\usepackage{amsmath}
\usepackage{epsfig}    
\usepackage{color}
\usepackage{slashed}
\usepackage{hhline}
\usepackage{ulem}
\usepackage{hyperref}

\def\be{\begin{equation}}
\def\ee{\end{equation}}
\newcommand{\bea}{\begin{eqnarray}}
\newcommand{\eea}{\end{eqnarray}}

\begin{document}

 \begin{flushright} {APCTP Pre2019- 005, KIAS-P19010}  \end{flushright}

\title{An inverse seesaw model with global $U(1)_H$ symmetry}

\author{Ujjal Kumar Dey}
\email{ujjal@apctp.org}
\affiliation{Asia Pacific Center for Theoretical Physics (APCTP) - Headquarters
San 31, Hyoja-dong, Nam-gu, Pohang 790-784, Korea}

\author{Takaaki Nomura}
\email{nomura@kias.re.kr}
\affiliation{School of Physics, KIAS, Seoul 02455, Republic of Korea}

\author{Hiroshi Okada}
\email{hiroshi.okada@apctp.org}
\affiliation{Asia Pacific Center for Theoretical Physics, Pohang 37673, Republic of Korea}
\affiliation{Department of Physics, Pohang University of Science and Technology, Pohang 37673, Republic of Korea}

\date{\today}

\begin{abstract}
We propose an inverse seesaw model based on hidden global symmetry $U(1)_H$ in which we realize tiny neutrino masses with rather natural manner taking into account relevant experimental bounds. The small Majorana mass for inverse seesaw mechanism is induced via small vacuum expectation value of a triplet scalar field whose Yukawa interactions  with standard model fermions are controlled by $U(1)_H$. We discuss the phenomenology of the exotic particles present in the model including the Goldstone boson coming from breaking of the global symmetry, and explore testability at the Large Hadron Collider experiments.
\end{abstract}

\maketitle

\section{Introduction}

%

Inverse seesaw mechanism~\cite{Mohapatra:1986bd, Wyler:1982dd} is one of the promising candidates to induce neutrino masses and their mixing that are typically understood by an additional symmetry beyond the standard model (SM).
The mechanism requires both chiralities of neutral fermions ($N_{L/R}$) that couple to the SM fermions with different manner, and their mass structures are also different. That is why an additional symmetry is needed~\cite{Cai:2018upp, Nomura:2018ktz, Nomura:2018cfu, Nomura:2018mwr}.
Furthermore, it demands hierarchies among mass parameters associated with neutral fermions including active neutrinos where especially Majorana mass of $N_L$ should be tiny.
Indeed, realizing such hierarchies in a natural way is rather challenging from model building perspective~\cite{Nomura:2018ktz, Nomura:2018cfu}.
In this paper, we propose an inverse seesaw model under a hidden {\it global} $U(1)_H$ symmetry~\cite{Okada:2014vla}~\footnote{There are several studies utilizing hidden {\it gauged} symmetries to explain neutrino masses~\cite{Nomura:2018ibs, Nomura:2017wxf, Nomura:2018kdi, Nomura:2018jkd, Yu:2016lof}.} in which we try to
realize natural hierarchies among neutral fermion mass matrix, taking advantage of experimental constraints of electroweak precision test to our model. 
In our model, we introduce $SU(2)_L$ triplet scalar and exotic lepton doublets with non-zero $U(1)_H$ charges and scalar singlets.
The Yukawa interaction among the triplet and exotic lepton doublet induces small Majorana mass term required for inverse seesaw mechanism due to small vacuum expectation value (VEV) of the triplet 
while Yukawa interaction between the SM lepton doublet and the scalar triplet is forbidden by the $U(1)_H$ symmetry. 
In addition, as a result of the global symmetry that accommodates a physical Goldstone boson (GB)~\cite{Weinberg:2013kea, Cheung:2017lpv}, our model can be well-tested at future colliders since the GB couples to the SM fermions through their kinetic terms.~\footnote{Even in the case of a gauged symmetry, GB is induced one introduces two or more two bosons that breaks the additional $U(1)$ symmetry; see, e.g., ref.~\cite{Nomura:2017wxf}.}

This letter is organized as follows.
In Sec. \ref{sec:model}, we introduce our model and formulate the scalar sector, charged-lepton sector, and neutral fermion sector, and briefly discuss electroweak precision test.
In the scalar sector, we show hierarchies among VEVs that connect to our desired hierarchies among neutral fermions. Using electroweak precision data, we  show the constraint on the VEV of an isospin triplet scalar at tree level, and this bound is directly related to the hierarchies of neutral fermion mass matrices.
Then we formulate the charged-lepton mass matrix that encapsulates with the mixing between SM charged-leptons and newly introduced heavier ones. From this we show that their mass hierarchies could be naturally realized,  while maintaining the experimental bound on the new heavier leptons.
Note that these hierarchies are also related to our neutral fermion mass matrix. In the last part of this section, we discuss the neutral fermion mass matrix and show how to realize these hierarchies. Also we discuss the bounds on non-unitarity effect originated from the inverse seesaw model.
In Sec. \ref{sec:pheno} we discuss the phenomenology of our model. We mainly consider the decays of the exotic charged leptons and they can leave their footprints in the multi-lepton signatures of the collider experiments, like LHC.
Finally we summarize our results and conclude.

\begin{table}[t!]
\begin{tabular}{|c||c|c|c|c||c|c|c|c|}\hline\hline  
& ~$L_L^a$~& ~$e_R^a$~& ~$L'^a_L$~& ~$L'^a_R$~& ~$H$~& ~$\Delta$~& ~$\varphi_1$~& ~$\varphi_2$~\\\hline\hline 
$SU(2)_L$   & $\bm{2}$  & $\bm{1}$  & $\bm{2}$  & $\bm{2}$ & $\bm{2}$   & $\bm{3}$ & $\bm{1}$ & $\bm{1}$   \\\hline 
$U(1)_Y$    & $-\frac12$  & $-1$ & $-\frac12$  & $-\frac12$  & $\frac12$ & {$1$}  &{$0$}  & ${0}$ \\\hline
$U(1)_H$    & $0$  & $0$ & $\ell$  & $2 \ell$  & $0$ & $2\ell$  & $\ell$  & $2\ell$ \\\hline
\end{tabular}
\caption{Charge assignments of the our lepton and scalar fields
under local $SU(2)_L\times U(1)_Y$ and global $U(1)_H$ symmetries, where the upper index $a$ is the number of family that runs over 1-3 and
all of them are singlet under $SU(3)_C$. }\label{tab:1}
\end{table}

\section{Model}
\label{sec:model}
In this section we formulate our model in which we introduce a global $U(1)_H$ symmetry.
The fermionic sector is augmented by three families of vector-like fermions $L'_{L/R}\equiv[N_{L/R},E_{L/R}]^T$ with charge $(\bm{2},-1/2)$ under the $SU(2)_L\times U(1)_Y$ gauge symmetry, while under the global $U(1)_{H}$ right- and left-handed ones are assigned charges $2\ell$ and $\ell$ respectively.
As for the scalar sector, we add an isospin triplet scalar $\Delta$ with hypercharge 1, and two singlet scalars $\varphi_{1,2}$ with zero hypercharge,
while ($2\ell,\ell,2\ell$) are respectively assigned to $(\Delta,\varphi_1,\varphi_2)$ under the global $U(1)_H$ symmetry. 
The SM-like Higgs field is denoted by $H$. The vacuum expectation values (VEVs) of ($H,\varphi_1,\varphi_2,\Delta$) are ($v/\sqrt2,v'_1/\sqrt2,v'_2/\sqrt2,v_\Delta/\sqrt2$), respectively.
The field content and the respective quantum number assignments are shown in Table~\ref{tab:1}. Note that the quark sector remains the same as that of SM.
The renormalizable Yukawa Lagrangian under these symmetries is then given by
\begin{align}
-{\cal L_\ell}
& =  y_{\ell_{aa}} \bar L^a_L H e^a_R  +  f_{ab} \bar L'^a_L L'^b_R \varphi^*_1 
+  g_{{L}_{ab}}  (\bar L'^c)^{a}_L  \tilde\Delta L'^b_L +y_{D_{ab}} \bar L^a_L L'^b_R \varphi^*_2
+ {\rm h.c.}, \label{Eq:yuk}
\end{align}
where the indices $a,b (=1-3)$ represent the number of families, $\tilde\Delta\equiv i\sigma_2\Delta^\dag$, and $y_\ell$ is assumed to be  diagonal matrix without loss of generality.
After spontaneous symmetry breaking, one finds the charged-lepton mass matrix $ m_\ell=y_\ell v/\sqrt2$.

\noindent \underline{\bf Scalar potential and VEVs}:\\
First of all, we define each scalar field as follows:
\begin{align}
H\equiv
\left[\begin{array}{c}
h^+  \\ 
\frac{v+h+ia}{\sqrt2} \\ 
\end{array}\right],\
\Delta\equiv \left[\begin{array}{cc}
 \frac{\delta^+}{\sqrt2} & \delta^{++}  \\ 
 \frac{v_\Delta+\delta_R+i\delta_I}{\sqrt2}   & - \frac{\delta^+}{\sqrt2} \\ 
\end{array}\right],\
\varphi_{1/2}\equiv\frac{v'_{1/2}+\varphi_{R_{1/2}}+i\varphi_{I_{1/2}}}{\sqrt2}.
\end{align}
The scalar potential in our model is given by,
\begin{align}
{\cal V} = & -\mu_h^2 H^\dagger H + M_\Delta^2 {\rm Tr}[\Delta^\dag\Delta] - \mu_1^2 |\varphi_1|^2 + M_2^2 |\varphi_2|^2
   -  (\mu_\varphi \varphi_2^* \varphi_1 \varphi_1 + h.c.) \nonumber \\
 & - [\lambda_0 H^T\tilde\Delta H \varphi_2 + h.c.] + \lambda_H (H^\dagger H)^2 + \lambda_{\varphi_1} |\varphi_1|^4 
 + \lambda_{\varphi_2} |\varphi_2|^4 + \lambda_{H \varphi_1} (H^\dagger H) |\varphi_1|^2 \nonumber \\
& + \lambda_{H \varphi_2} (H^\dagger H) |\varphi_2|^2 + \lambda_{\varphi_1 \varphi_2}  |\varphi_1|^2 |\varphi_2|^2 
+ {\cal V}_{\rm trivial},
\end{align}
where ${\cal V}_{\rm trivial}$ indicates trivial quartic terms containing scalar triplet.  For simplicity we assume all the couplings to be real. The quartic coupling $\lambda_{0}$ plays a role in reducing the scale of VEV of $\Delta$ to $\mathcal{O}(1)$ GeV. 
Applying the minimization condition $\partial {\cal V}/\partial v= \partial {\cal V}/\partial v'_{1,2}= \partial {\cal V}/\partial v_\Delta = 0$, we obtain the VEVs approximately as 
\begin{align}
v \simeq \sqrt{\frac{\mu_h^2}{\lambda_H}}, \quad v'_1 \simeq \sqrt{\frac{\mu_1^2}{\lambda_{\varphi_1}}}, \quad v'_2 \simeq \frac{\sqrt{2} \mu_\varphi v'^2_1}{2M_2^2 +  \lambda_{\varphi_1 \varphi_2} v'^2_1}, \quad v_\Delta \simeq \frac{\lambda_0 v^2 v'_2}{4 M_\Delta^2},
\end{align}
where we assume $v'_2 \sim v_\Delta \ll \{ v'_1, v\}$ and $\lambda_{H\varphi_1}$ to be small.
The above hierarchy of VEVs are motivated from the mass hierarchy in neutral fermions as we discuss below. 
{To obtain the VEV hierarchy and electroweak vacuum consistently we need to choose appropriate parameters.
When we impose $v'_1 \gtrsim 5 \times 10^6$ GeV as required by constraints from massless Goldstone boson interactions as discussed below, and taking $v'_2 \sim v_\Delta = \mathcal{O}(1)$ GeV and $M_2^2 \sim \lambda_{\varphi_1 \varphi_2} v'^2_1 \sim M_\Delta^2 \sim v^2$,
we should choose $\mu_\varphi \lesssim 10^{-8}$ GeV and $\lambda_{\varphi_1 \varphi_2} \lesssim 10^{-8}$.   
Furthermore $\lambda_{H \varphi_1}$ should be very small as $\mathcal{O}(10^{-8})$ or less when $v'_1 \gtrsim 5 \times 10^6$ GeV 
since $\sqrt{\lambda_{H \varphi_1}} v'_1 \lesssim \mathcal{O}(100)$ GeV in order to realize electroweak vacuum.
Also one loop contribution to $\lambda_{H \varphi_1}$ is roughly obtained {as $\sim \lambda_{H \varphi_2} \lambda_{\varphi_1 \varphi_2}/(4 \pi)^2$, 
and the condition $\lambda_{H \varphi_1} \lesssim 10^{-8}$ requires $\lambda_{H \varphi_2} \lambda_{\varphi_1 \varphi_2} \lesssim 10^{-6}$.} 
}

Notice that smallness of $\mu_{\varphi}$ would be natural in the 't Hooft sense since we can restore a ``lepton number" symmetry for $\mu_\varphi \to 0$ limit if the quantum number is assigned as 
$1$ for fileds $\{ L_L, e_R, L'_R, \varphi_1 \}$ and $0$ for $\{ L'_L, \varphi_2, \Delta\}$. 

Since we assume small $\lambda_{H \varphi_1}$ coupling, the mass matrix for CP-even scalars from the SM singlets $\varphi_{1,2}$ is derived approximately 
\begin{equation}
\frac{1}{2} \left( \begin{array}{c} \varphi_{R_1} \\ \varphi_{R_2} \end{array} \right)^T 
\left( \begin{array}{cc} 2 \lambda_{\varphi_1} v'^2_1 & \lambda_{\varphi_1 \varphi_2} v'_1 v'_2 - \sqrt{2} \mu_\varphi v'_1 \\ \lambda_{\varphi_1 \varphi_2} v'_1 v'_2 - \sqrt{2} \mu_\varphi v'_1 & M_2^2 + \frac{1}{2} \lambda_{\varphi_1 \varphi_2} v'^2_1  \end{array} \right)
\left( \begin{array}{c} \varphi_{R_1} \\ \varphi_{R_2} \end{array} \right).
\end{equation}
Note that mixing between $\varphi_{R_1}$ and $\varphi_{R_2}$ would be small since off-diagonal elements are suppressed by $v'_2$ and $\mu_\varphi$ unless values of diagonal elements are not too close.
Thus we approximate $\varphi_{R_1}$ and $\varphi_{R_2}$ as mass eigenstates and corresponding mass eigenvalues are 
\begin{equation}
m_{\varphi_{R_1}} \simeq \sqrt{2 \lambda_{\varphi_1}} v'_1, \quad m_{\varphi_{R_2}} \simeq \sqrt{M_2^2 + \frac{1}{2} \lambda_{\varphi_1 \varphi_2} v'^2_1}.
\end{equation}  

In CP-odd scalar sector from SM singlet $\varphi_{1,2}$, we have one massless Goldstone boson associated with breaking of global $U(1)_H$ symmetry.
Applying our assumption of $v'_2 \ll v'_1$, we can approximately identify $\varphi_{I_1}$ as Goldstone boson while $\varphi_{I_2}$ is massive scalar boson with mass value $m_{\varphi_{I_2}} \simeq \sqrt{M_2^2 + \lambda_{\varphi_1 \varphi_2} v'^2_1/2}$. 
Thus the $\varphi_{R_2}$ and $\varphi_{I_2}$ have approximately degenerate masses.
{In our following analysis, we consider massive scalars from $\varphi_1$ and $\varphi_2$ have electroweak scale masses expecting collider phenomenology of these scalar bosons. We then impose that $\lambda_{\varphi_1} \lesssim 10^{-8}$ since $v'_1$ should be larger than $\sim 5 \times 10^6$ GeV from constraints of Goldstone boson having axion-like interaction as discussed below. In addition $\lambda_{\varphi_1 \varphi_2} \lesssim 10^{-3}$ is required since $\varphi_2$ loop correction to $\lambda_{\varphi_1}$ is roughly estimated to be $\sim \lambda_{\varphi_1 \varphi_2}^2 /(4 \pi)^2$. Clearly with these choices of quartic couplings one can have the massive scalars in the electroweak scale relevant for the collider phenomenology.
}

The SM Higgs and scalar triplet sector are mostly same as the usual scalar triplet model since we assume the mixing among SM singlet scalar to be sufficiently small. 
The mass scale of scalar bosons in the triplet is given by $M_\Delta$.

\noindent \underline{\bf $\rho$ parameter}:\\
The VEV of $\Delta$ is restricted by the $\rho$-parameter at tree level  that is given by~\cite{Kanemura:2012rs} 
\begin{align}
\rho\approx \frac{v^2+ 2 v_\Delta^2}{v^2+ 4 v_\Delta^2},
\end{align}
where the experimental value is given by $\rho=1.0004^{+0.0003}_{-0.0004}$ at $2\sigma$ confidence level~\cite{pdg}.
On the other hand, we have $v_{SM}=\sqrt{v^2+ 2 v_\Delta^2}\approx$246 GeV. Therefore the upper bound on $v_\Delta$ is of the order $\mathcal{O}(1)$ GeV~\footnote{Theoretical origins are studied by refs.~\cite{Kanemura:2012rj, Nomura:2016dnf, Okada:2015nca}.}.

\noindent \underline{\bf Charged-lepton sector}:\\
The charged-lepton mass matrix consists of the component of the SM mass matrix and heavier one,
after the spontaneous symmetry breaking. We define the mass matrix $M$ to be $fv'_1/\sqrt2$.
Furthermore, we assume the mass matrices $m_D$ and $M$ to be diagonal for simplicity.
Then, for each generation indicated by "$a$", we can write the charged-lepton fermion mass matrix as
\begin{align}
& \left( \begin{array}{c} \bar e_L^a \\ \bar E_L^a \end{array} \right)^T
M_{E_a}
\left( \begin{array}{c}  e_R^a \\  E_R^a \end{array} \right)
=
\left( \begin{array}{c} \bar e_L^a \\ \bar E_L^a \end{array} \right)^T
\left[\begin{array}{cc}
 m_{\ell_a} & m_{D_a}  \\ 
0  & M_a  \\ 
\end{array}\right]
\left( \begin{array}{c}  e_R^a \\  E_R^a \end{array} \right),\\
& M_{E_a} M_{E_a}^\dag
=
\left[\begin{array}{cc}
 m_{\ell_a}^2  + m_{D_a}^2 & m_{D_a} M_a  \\ 
m_{D_a} M_a & M_a^2  \\ 
\end{array}\right], \quad 
M_{E_a}^\dag M_{E_a}
=
\left[\begin{array}{cc}
 m_{\ell_a}^2 & m_{D_a}  m_{\ell_a}  \\ 
m_{D_a}  m_{\ell_a} & M_a^2   + m_{D_a}^2   \\ 
\end{array}\right].
\end{align}
The mass matrix is diagonalized by the transformation $(e^a_{L(R)}, E^a_{L(R)}) \to V_{L_a(R_a)}^\dag (e^a_{L(R)}, E^a_{L(R)})$.
Thus we can obtain diagonalization matrices $V_{L_a}$ and $V_{R_a}$ which respectively diagonalize $M_{E_a} M_{E_a}^\dag$ and $M_{E_a}^\dag M_{E_a}$ as $V_{L_a} M_{E_a} M_{E_a}^\dag V_L^\dag =V_{R_a} M_{E_a}^\dag M_{E_a} V_{R_a}^\dag={\rm diag}(m^2_{e_a}, M^2_{E_a})$, such that 
\begin{align}
& V_{L_a} = \left( \begin{array}{cc} \cos \theta_{L_a} & - \sin \theta_{L_a} \\ \sin \theta_{L_a} & \cos \theta_{L_a}  \end{array} \right), \quad 
V_{R_a} = \left( \begin{array}{cc} \cos \theta_{R_a} & - \sin \theta_{R_a} \\ \sin \theta_{R_a} & \cos \theta_{R_a}  \end{array} \right),  \\
& \tan 2 \theta_{L_a} = \frac{2 m_{D_a} M_a}{M_a^2 - m_{\ell_a}^2 - m_{D_a}^2} \simeq \frac{2 m_{D_a}}{M_a}, \quad 
\tan 2 \theta_{R_a} = \frac{2 m_{D_a} m_{\ell_a}}{M_a^2 + m_{D_a}^2 - m_{\ell_a}^2 } \simeq \frac{2 m_{D_a} m_{\ell_a}}{M_a^2},
\end{align}
where we have assumed $m_{\ell_a},m_{D_a} \ll M_a$.
 Then the mass eigenvalues for $e^a$ and $E^a$ are simply given by $m_{\ell_a}$ and $M_a$. 
Also their mixing angles $\theta_{R_a}$ and $\theta_{L_a}$ are very small and satisfy $\theta_{R_a} \ll \theta_{L_a}$, since  the lower bound on the mass of the heavier leptons is about 100 GeV~\cite{pdg} that is suggested by the current experimental data at LHC and LEP.
This is because hierarchies in mass parameter $m_{\ell_a}, m_{D_a} \ll M_a$ is comparatively natural.
Note also that the constraints from electroweak precision measurements ($S,T,U$-parameters) can be avoided when the components in exotic lepton doublet have degenerate mass. In fact we consider such a case where mass of $E$ and $N$ are dominantly given by Dirac mass $M$.

\noindent \underline{\bf Neutrino sector}:\\
After the spontaneous symmetry breaking, neutral fermion mass matrix in the basis of $(\nu^c_L,N_R,N_L^c)^T$ is given by
\begin{align}
M_N
&=
\left[\begin{array}{ccc}
0 & m_D & 0  \\ 
m_D & 0 & M \\ 
0  & M & \mu^* \\ 
\end{array}\right],
\end{align}
where $m_D,M$ are diagonal while $\mu\equiv g_L v_\Delta/\sqrt2$ is symmetric $3\times 3$ mass matrix.
Then the active neutrino mass matrix can be given as
\begin{align}
m_\nu\approx  \mu^* \left(\frac{m_D}{M}\right)^2.
\end{align}
Once we fix $m_D/M\sim{\cal O}(0.01)$, then $\mu\sim {\cal O}(10^{-7})$ GeV in order to satisfy the observed neutrino mass squared differences.
It suggests that $g_L \sim {\cal O}(10^{-7})$ when we fix $v_\Delta\sim {\cal O}(1)$ GeV\footnote{These hierarchies could be explained by several mechanisms such as radiative models~\cite{Dev:2012sg, Dev:2012bd, Das:2017ski} and effective models with higher order terms \cite{Okada:2012np}.}.
The neutrino mass matrix is diagonalized by unitary matrix $U_{MNS}$; $D_\nu= U_{MNS}^T m_\nu U_{MNS}$, where $D_\nu\equiv {\rm diag}(m_1,m_2,m_3)$.
%
%
Constraint from non-unitarity can simply be obtained by considering the hermitian matrix $F\equiv  M_{}^{-1} m_D$.
Combining several experimental results~\cite{Fernandez-Martinez:2016lgt},
the upper bounds are given by~\cite{Agostinho:2017wfs}:
\begin{align}
|FF^\dag|\le  
\left[\begin{array}{ccc} 
2.5\times 10^{-3} & 2.4\times 10^{-5}  & 2.7\times 10^{-3}  \\
2.4\times 10^{-5}  & 4.0\times 10^{-4}  & 1.2\times 10^{-3}  \\
2.7\times 10^{-3}  & 1.2\times 10^{-3}  & 5.6\times 10^{-3} \\
 \end{array}\right].
\end{align} 
Since $F$ is assumed to be diagonal, the stringent constraint originates from 2-2 component of $|FF^\dag|$.
It suggests that $|F|\sim m_D/M\lesssim 10^{-2}$ that is always safe in our model thanks to the small VEV $v'_2$.

\section{Phenomenology}
\label{sec:pheno}
In this section, we discuss the phenomenology of the model focusing on the decay and production of exotic particles. 
One specific property of our model is the existence of physical Goldstone boson (GB) as a result of global $U(1)_H$ symmetry breaking~\footnote{Physical GB would obtain tiny mass due to breaking of global $U(1)$ symmetry by gravitational effects~\cite{Kallosh:1995hi}. In our analysis we simply consider it as a massless particle.}.
Here we first derive interactions associated with the Goldstone boson which is denoted by $\alpha_G$ identified as   
$\varphi_1 = e^{i \frac{\alpha_G}{v'_1}} (v'_1 + \varphi_{R_1})/\sqrt{2}$ in the limit of $v'_2 \ll v'_1$. 
Then fields with $U(1)_H$ charge are rewritten as $\varphi_2 \to e^{i \frac{2 \alpha_G}{v'_1}} \varphi_2$, $\Delta \to e^{i \frac{2\alpha_G}{v'_1}} \Delta$, 
$L'^a_L \to e^{i \frac{\alpha_G}{v'_1}} L'^a_L$ and $L'^a_R \to e^{i \frac{2\alpha_G}{v'_1}} L'^a_R$.
Then $\varphi_{1}$ and $\varphi_2$ interact with $\alpha_G$ as follows
\begin{align}
{\cal L} \supset & \frac{1}{v'_1} \varphi_{R_1} \partial_\mu \alpha_G \partial^\mu \alpha_G + \frac{1}{2 v'^2_1} \varphi_{R_1} \varphi_{R_1} \partial_\mu \alpha_G \partial^\mu \alpha_G
+ i \frac{2}{v'_1} \partial_\mu \alpha_G (\partial^\mu \varphi_{R_2} \varphi_{I_2} - \varphi_{R_2} \partial^\mu \varphi_{I_2} )  \nonumber \\
& + \frac{2 v'_2}{v'^2_1} \varphi_{R_2} \partial_\mu \alpha_G \partial^\mu \alpha_G + \frac{1}{ v'^2_1} (\varphi_{R_2} \varphi_{R_2} + \varphi_{I_2} \varphi_{I_2}) \partial_\mu \alpha_G \partial^\mu \alpha_G.
\end{align}
The covariant derivative of $\Delta$ is rewritten including Goldstone boson as, 
\begin{align}
D_{\mu} \Delta = & \partial_{\mu} \Delta + i \frac{2}{v'_1} \partial_\mu \alpha_G \Delta -i \frac{g}{\sqrt{2}} \left( W_{\mu}^+ [T^+,\Delta]+W_{\mu}^-[T^-,\Delta] \right) \nonumber \\ 
& -i \frac{g}{\cos \theta_W} Z_{\mu} \left( \left[ \frac{\sigma_3}{2},\Delta \right]- \sin^2 \theta_W \hat Q \Delta \right) -ie A_{\mu} \hat Q \Delta,
\label{eq:Eint}
\end{align}
where $g$ is $SU(2)_L$ gauge coupling, $T^\pm = (\sigma_1 \pm i\sigma_2)/2$, $\theta_W$ is Weinberg angle, and $\hat Q$ is electric charge operator acting on each component of the multiplet.
Then we can obtain interactions of $\alpha_G$ and $\Delta$ from kinetic term Tr$[(D_\mu \Delta)^\dagger(D^\mu \Delta)]$.
The triplet Higgs decay modes with $\alpha_G$ are suppressed by $v_\Delta/v'_1$ factor so that components in $\Delta$ decay via gauge or scalar potential interactions.  
Also exotic lepton $L'^a$ interaction with $\alpha_G$ is derived from kinetic term as 
\begin{align}
{\cal L} \supset & - \bar L'^a \gamma^\mu \left( \frac{1}{v'_1} \partial_\mu \alpha_G P_L +  \frac{2}{ v'_1} \partial_\mu \alpha_G P_R  \right) L'^a \nonumber \\
& +  \bar L'^a \gamma^\mu \left( \frac{g}{\sqrt{2}} (W^+_\mu T^+ + W^-_\mu T^-) + \frac{g}{\cos \theta_W} \left(\frac{\sigma_3}{2} - \sin^2 \theta_W \hat Q \right) Z_\mu + e \hat Q A_\mu \right) L'^a.
\end{align}
Here we briefly discuss constraints on our physical GB.
The GB can have axion-like photon coupling since assignment of $U(1)_H$ charge for extra leptons $L'$ depends on its chirality.
Then we roughly obtain the effective interaction as~\cite{Kim:2008hd}
\begin{equation}
\mathcal{L}_{\alpha_G \gamma \gamma} \sim \frac{e^2}{32 \pi^2 v'_1} \alpha_G{F_{\mu\nu} \tilde F^{\mu\nu}}, 
\end{equation}   
where $F_{\mu \nu}$ is photon field strength.
Then cooling of stars provide a constraint for this coupling such that~\cite{Ayala:2014pea,Sedrakian:2015krq} 
\begin{equation}
\frac{e^2}{32 \pi^2 v'_1} \lesssim 6.6 \times 10^{-11} \ {\rm GeV}^{-1} \rightarrow v'_1 \gtrsim 4.6 \times 10^6 \ {\rm GeV}.
\end{equation}
We thus consider $v'_1$ to be very high scale as $5 \times 10^6$ GeV. 
In addition, experiments which search for new force mediated by axion-like particles provide another constraint~\cite{Sushkov:2011zz,Arvanitaki:2014dfa} but it is less severe compared to the star cooling one.
If we take large $v'_1$ the existence of our GB does not cause serious problem in cosmology since it does not couple to the SM particles via some direct coupling except to the Higgs boson whose couplings are well suppressed by $1/v'_1$ factor and controlled by the parameters in the potential which we assume to be small. 
Thus the GB decouples from thermal bath in sufficiently early Universe since all hidden particles are heavier than $\mathcal{O}(100)$ GeV scale, and it almost does not affect number of relativistic degrees of freedom in  the current Universe.
Even if the GB interacts with SM particle via Higgs portal, cosmological constraint can be avoided when the GB decouples from the thermal bath at the scale larger than muon mass scale~\cite{Weinberg:2013kea}.


\subsection{Decay of exotic particles}
In this subsection, we discuss the decays of exotic particles in the model.
At first, we write the Lagrangian relevant to decay of exotic charged lepton $E_a$ as follows 
\begin{align}
{\cal L} \supset & \frac{1}{v'_1} \frac{m_{D_a}}{M_a} \bar \ell^a \gamma^\mu P_L E^a \partial_\mu \alpha_G +  f_{aa} \frac{m_{D_a}}{M_a} \bar \ell^a P_R E^a \varphi_{R_1}
 + \frac{y_{D_{aa}}}{\sqrt{2}} \bar \ell^a P_R E^a (\varphi_{R_2} + i \varphi_{I_2}) \nonumber \\
& + \frac{y_{D_{aa}}}{\sqrt{2}} \frac{m_{D_a} m_{\ell_a}}{M_a^2} \bar \ell^a P_R \ell^a (\varphi_{R_2} + i \varphi_{I_2})  + h.c. \, , 
\end{align}
where we have omitted subdominant terms.
The partial decay widths are computed as 
\begin{align}
& \Gamma_{E^{a-} \to \ell^{a-} \varphi_{R_1}} \simeq \frac{f_{aa}^2}{64 \pi M_a} \frac{y^2_{D_{aa}} v'^2_2}{M^2_a} (M_a^2 - m_{\varphi_2}^2), \\
& \Gamma_{E^{a-} \to \ell^{a-} \varphi_{R_2}} = \Gamma_{E^a \to \ell^a \varphi_{I_2}} \simeq \frac{y_{D_{aa}}^2}{64 \pi M_a} (M_a^2 - m_{\varphi_2}^2), \\
& \Gamma_{E^{a-} \to \ell^{a-} \alpha_G} \simeq \frac{y_{D_{aa}}^2 v'^2_2}{32 \pi v'^2_1} M_a,
\end{align}
where we have used $m_{D_a} = y_{D_{aa}} v'_2/\sqrt{2}$. 
Thus $E^{a\pm}$ dominantly decay into $\ell^{a\pm} \varphi_{R_2(I_2)}$ since the other modes are suppressed by small VEV $v'_2$. 
Then we also estimate partial decay widths of $\varphi_{R_2}(\varphi_{I_2})$ as,
\begin{align}
& \Gamma_{\varphi_{R_2} \to \alpha_G \alpha_G} \simeq \frac{m_{\varphi_2}^3 v'^2_2}{8 \pi v'^4_1}, \\
& \Gamma_{\varphi_{R_2} \to \ell^{a-} \ell^{a+}} = \Gamma_{\varphi_{I_2} \to \ell^a \bar \ell^a} \simeq \frac{y_{D_{aa}}^4}{32 \pi} \frac{v'^2_2 m_{\ell^a}^2}{M_a^4} m_{\varphi_2}. 
\end{align}
%
Thus we find that $\varphi_{R_2}$ and $\varphi_{I_2}$ dominantly decays into $\ell^{a+} \ell^{a-}$ when we take $v'_1 = 5 \times 10^{6}$ GeV.
Therefore dominant decay chains of $E^{a -}$ are $E^{a -} \to \ell^{a-} \varphi_{I_2} \to \ell^{a-} \ell^{b-} \ell^{b+}$ and $E^{a -} \to \ell^{a-} \varphi_{R_2} \to \ell^{a-} \ell^{b-} \ell^{b+}$ with
branching ratio (BR) of $0.5$ for both modes.
Notice also that here we took coupling among $\varphi_R$ and SM Higgs to be zero but $\varphi_R$ can decay SM particles through such a coupling. 

\subsection{Collider physics}
\begin{figure}[tb]
\begin{center}
\includegraphics[width=7.0cm]{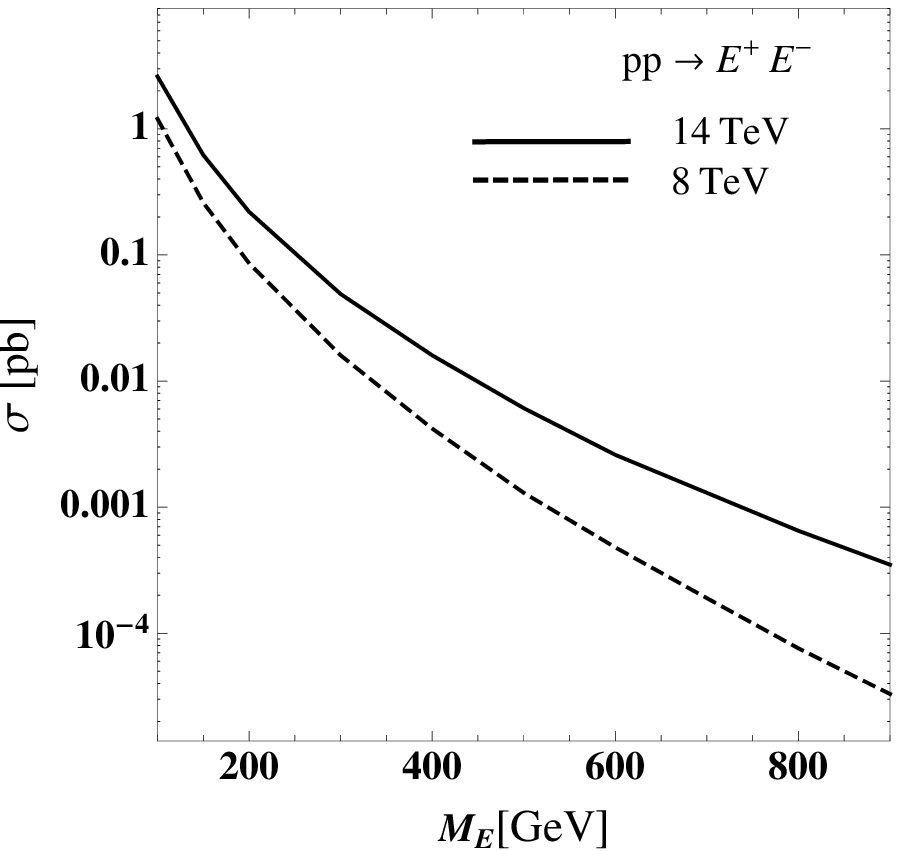}
\includegraphics[width=7.0cm]{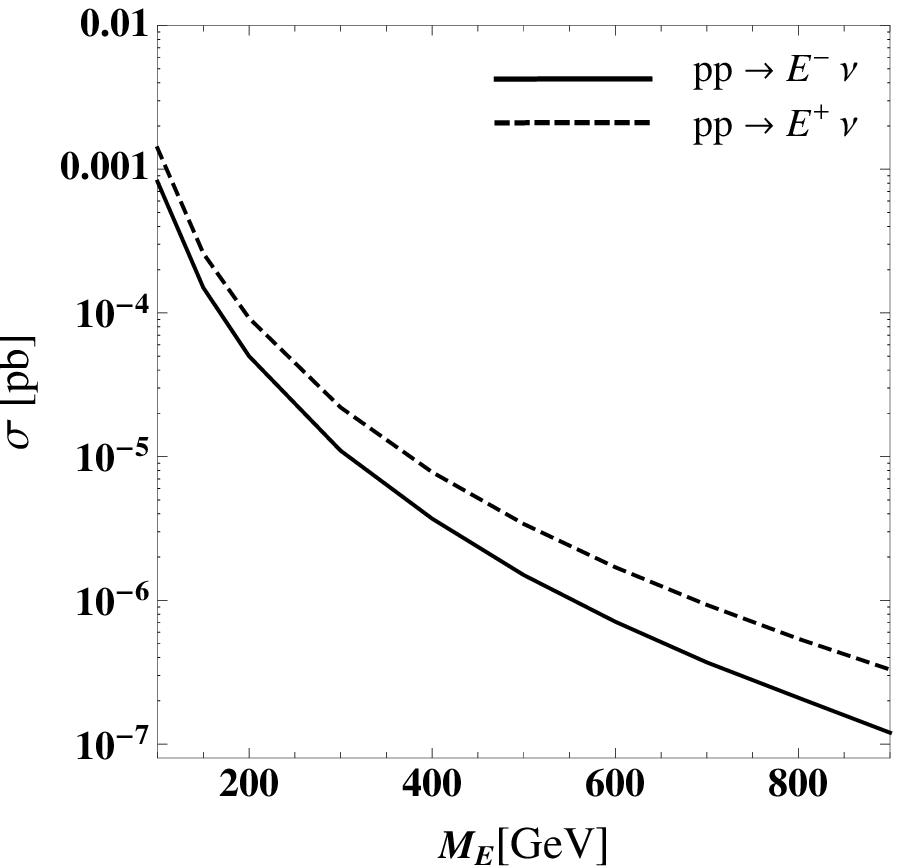}
\caption{Left: $E^+E^-$ pair production cross section as a function of its mass. Right: $E^\pm$ single production cross section as a function of its mass where we fix $m_D/M = 0.01$.}
\label{fig:CX}
\end{center}\end{figure}

In this subsection we discuss collider signatures of our model.
The exotic charged scalar bosons from Higgs triplet can be produced by electroweak production and they dominantly decay into gauge bosons as the triplet components have degenerate mass.
This phenomenology is the same as scalar triplet model with relatively large triplet VEV case ($v_\Delta \sim 1$ GeV) 
where charged scalars in the triplet $\{\delta^\pm, \delta^{\pm \pm} \}$ decays into SM gauge bosons.
Phenomenology of such case can be found, for example, in refs.~\cite{Perez:2008ha, Chiang:2012dk, Du:2018eaw, Ghosh:2017pxl, Kanemura:2014ipa}.

Then we focus on exotic charged lepton production at the LHC. 
The exotic charged lepton pair can be produced by $pp \to Z/\gamma \to  E^{a+} E^{a-}$ via gauge interaction depicted in Eq.(\ref{eq:Eint}). 
Furthermore single production process can be induced through mixing between $E^a$ and SM charged leptons.
We obtain relevant interaction by 
\begin{equation}
{\cal L} \supset \frac{g}{\sqrt{2}} \frac{m_{D_a}}{M_a} \bar E^a \gamma^\mu P_L \nu^a W_\mu^- + h.c. \, ,
\end{equation}
where mixing effect in neutral current is canceled. 
Thus $E^a$ can be singly produced as $pp \to W^{-(+)} \to E^{a-} \bar \nu^a (E^{a+} \nu^a)$ at the LHC.
For these processes cross sections are estimated by using \texttt{CalcHEP}~\cite{Belyaev:2012qa} with the CTEQ6 parton distribution functions (PDFs)~\cite{Nadolsky:2008zw}.
We show pair and single production cross sections as a function of exotic lepton mass in left and right panels in Fig.~\ref{fig:CX}, where we fixed $m_{D_a}/M_a = 0.01$ in our calculation and only the lightest exotic charged lepton $E^\pm$ is considered.
Then we find that single production cross section is much smaller than pair production one due to the suppression by mixing factor $m_{D_a}/M_a$.
The pair production cross section is larger than $\sim 1$ fb when exotic lepton mass is $M_{a} \lesssim 700$ GeV for $\sqrt{s} = 14$ TeV.
Since the exotic charged lepton decays into multi-lepton final state, the cross section can be constrained by multi-lepton search at the LHC
where inclusive search indicates cross section producing more than three electron/muon is required to be $\sigma \cdot BR \lesssim 1$ fb at LHC 8 TeV~\cite{Aad:2014hja}.
We thus require our exotic charged lepton mass to be $M_a \gtrsim 500$ GeV. 
%

\begin{figure}[tb]
\begin{center}
\includegraphics[width=7.0cm]{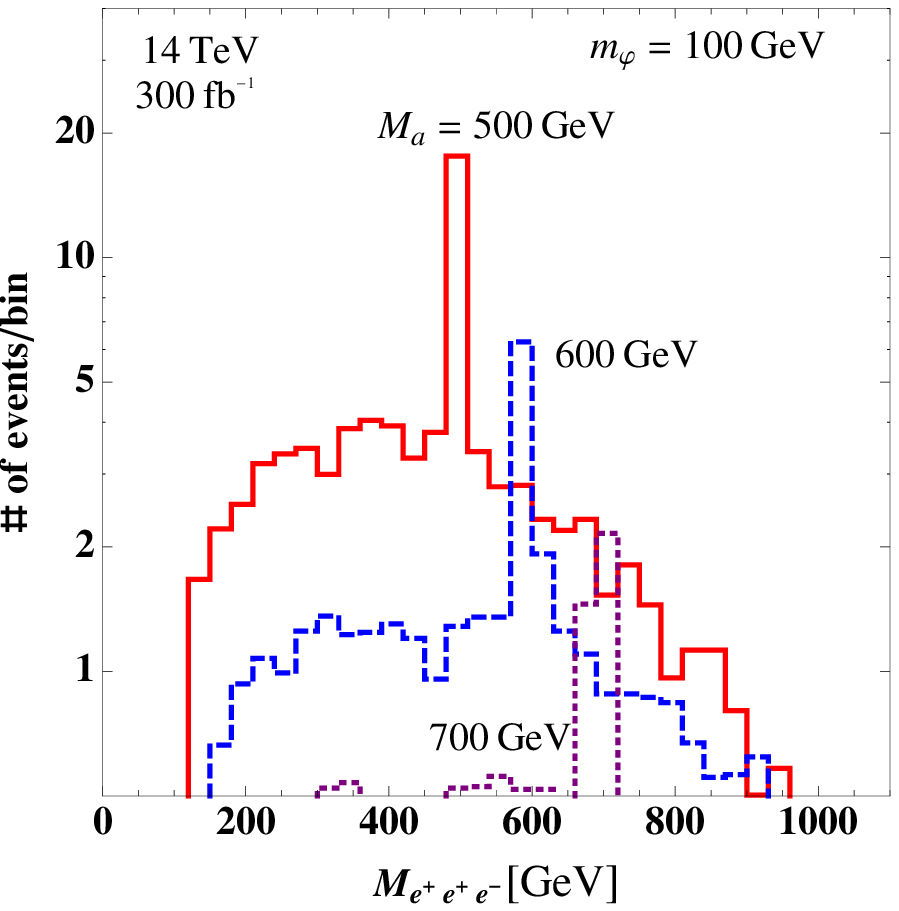}~~~~
\includegraphics[width=7.0cm]{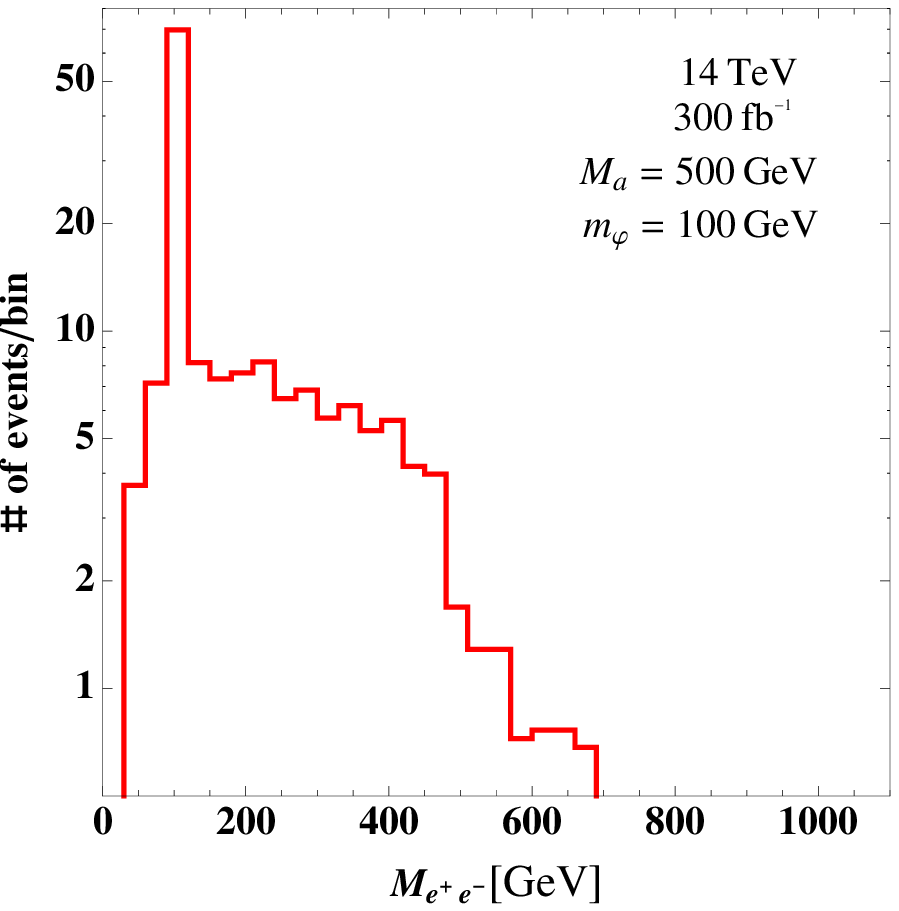}
\caption{Left:Number of events for invariant mass for $e^+ e^+ e^-$ with $m_\varphi \equiv m_{\varphi_{I_2,R_2}} = 100$ GeV and integrated luminosity 300 fb$^{-1}$ at the LHC 14 TeV 
Right: Number of events for invariant mass for $e^+ e^-$. }
\label{fig:event}
\end{center}\end{figure}

%
{One specific signal from $E^{+} E^{-}$ pair production is multi-lepton process given by $E^{+}  E^{-} \to \ell^+ \ell^- \varphi_{I_2,R_2} \varphi_{I_2,R_2} \to \ell^+ \ell^-  \ell^+  \ell^-  \ell^+  \ell^- $. Notice that 
we require scalar bosons $\varphi_{I_2,R_2}$ to have masses of $\mathcal{O}(100)$ GeV so that the decay process is dominant; if these scalar bosons are much heavier, the dominant decay mode of $E^\pm$ is $W^\pm \nu$ mode through mixing between heavy extra lepton and the SM leptons.}
Here, signal events are generated by employing the event generator {\tt MADGRAPH/MADEVENT\,5}~\cite{Alwall:2014hca}, where the necessary Feynman rules and relevant parameters of the model are implemented by use of {\tt FeynRules 2.0} \cite{Alloul:2013bka} and the {\tt NNPDF23LO1} PDF~\cite{Deans:2013mha} is adopted. 
Then the  {\tt PYTHIA\,6}~\cite{Ref:Pythia}  is applied to deal with hadronization effects,  the  initial-state radiation (ISR) and final-state radiation (FSR) effects, 
and the generated events are also run though the {\tt PGS\,4} for detector level simulation~\cite{Ref:PGS}. 
Then we select events with 6 charged leptons and impose cuts for lepton transverse momentum as $p_T(\ell) > 15$ GeV.
In left panel of Fig.~\ref{fig:event}, we show number of events distribution for invariant mass for $e^+ e^+ e^-$ assuming $E^\pm$ and $\varphi_{I_2,R_2}$ dominantly decay into mode including electron and $m_{\varphi_{I_2,R_2}} = 100$ GeV with integrated luminosity 300 fb$^{-1}$ at the LHC 14 TeV~\footnote{Invariant mass for $e^- e^- e^+$ provides similar distribution.}.
We see the clear peak at the exotic lepton mass and distributions around the peak as we can not always choose $e^+ e^+ e^-$ combination coming from $E^+$ decay chain.
In addition we show distribution for invariant mass for $e^+ e^-$ in right panel of Fig.~\ref{fig:event} which shows the peak at the mass of $\varphi_{I_2, R_2}$.
The signal is clean and number background (BG) events can be suppressed requiring multiple charged lepton final state.
In fact number of BG events is negligibly small requiring 6 charged lepton in final state where the largest cross section from SM processes is 
$\sigma (pp \to Z Z Z, Z \to \ell^+ \ell^-)$ and it is around $10^{-3}$ fb magnitude. 
Thus the signal can be tested well at the future LHC experiments if exotic lepton mass is less than $1$ TeV.
More detailed simulation including sophisticated cut analysis is beyond the scope of this paper and it is left for future work.  


\section{Summary and Conclusions}
We have constructed an inverse seesaw model based on $U(1)_H$ global symmetry in which we introduced exotic lepton doublets $L'$ and new scalar fields including scalar triplet with $U(1)_H$ charge.
The exotic lepton doublets are vector-like under gauge symmetry but chiral under $U(1)_H$ and Majorana mass of neutral components are zero before $U(1)_H$ symmetry breaking.
Then $U(1)_H$ is spontaneously broken by VEVs of scalar fields and only neutral component of $L'_L$ obtains Majorana mass term via triplet VEV due to appropriate $U(1)_H$ charge assignment.
We thus realize neutral fermion mass matrix which has the structure similar to inverse seesaw mechanism.
%
Our advantage of this model is to realize the neutrino mass matrix with natural mass parameters in the framework of inverse seesaw scenario. Therefore, the lepton number is broken by exotic fermions and its violation is restricted by the triplet VEV, that is  of the order 1 GeV, originated from the constraint of oblique parameters. Also $m_D/M$, which is also proportional to the neutrino mass matrix, is naturally suppressed, when we appropriately assign the lepton number to the fields. This is because $m_D$ is proportional to $\mu_\varphi\sim\cal O$(1) GeV, which is the source of lepton number violation, that is expected to be small due to the 't Hooft sense, while $M$ should be greater than 100 GeV which is required by the current experimental data such as LHC and LEP. We thus have naturally achieved that the neutrino mass matrix be less than the order of $10^{-4}$ GeV.
%

%
We have also discussed the phenomenology of the model focusing on decay and production processes of the exotic particles.
The decay widths of exotic charged leptons $E^\pm$ are estimated taking into account interactions with Goldstone boson from $U(1)_H$ symmetry breaking.
We have found $E^\pm$ will provide multi-lepton final states from decay chain.
Then $E^\pm$ production cross section has been estimated which is induced by electroweak interactions.
Testable number of multi-lepton events could be obtained if mass of $E^\pm$ is less than $\mathcal{O}(1)$ TeV scale. 

\section*{Acknowledgments}
This research is supported by the Ministry of Science, ICT \& Future Planning of Korea, the Pohang City Government, and the Gyeongsangbuk-do Provincial Government (U. K. D. and H. O.).
H. O. is sincerely grateful to KIAS and all the members.


\begin{thebibliography}{99}


\bibitem{Mohapatra:1986bd} 
  R.~N.~Mohapatra and J.~W.~F.~Valle,
  Phys.\ Rev.\ D {\bf 34}, 1642 (1986).
  

\bibitem{Wyler:1982dd} 
  D.~Wyler and L.~Wolfenstein,
  Nucl.\ Phys.\ B {\bf 218}, 205 (1983).
  
\bibitem{Cai:2018upp} 
  H.~Cai, T.~Nomura and H.~Okada,
  arXiv:1812.01240 [hep-ph].
  
\bibitem{Nomura:2018ktz} 
  T.~Nomura and H.~Okada,
  arXiv:1809.06039 [hep-ph].
  
\bibitem{Nomura:2018cfu} 
  T.~Nomura and H.~Okada,
  arXiv:1807.04555 [hep-ph].
  
\bibitem{Nomura:2018mwr} 
  T.~Nomura and H.~Okada,
  LHEP {\bf 1}, no. 2, 10 (2018)
  [arXiv:1806.01714 [hep-ph]].
  
  
\bibitem{Okada:2014vla} 
  H.~Okada,
  arXiv:1404.0280 [hep-ph].
  
  
\bibitem{Nomura:2018ibs} 
  T.~Nomura and H.~Okada,
  arXiv:1806.07182 [hep-ph].
   
\bibitem{Nomura:2017wxf} 
  T.~Nomura and H.~Okada,
  Phys.\ Rev.\ D {\bf 97}, no. 7, 075038 (2018)
  [arXiv:1709.06406 [hep-ph]].
   
\bibitem{Nomura:2018kdi} 
  T.~Nomura and H.~Okada,
  arXiv:1812.08473 [hep-ph].
   
\bibitem{Nomura:2018jkd} 
  T.~Nomura, H.~Okada and P.~Wu,
  JCAP {\bf 1805}, no. 05, 053 (2018)
  [arXiv:1801.04729 [hep-ph]].
  
\bibitem{Yu:2016lof} 
  J.~H.~Yu,
  Phys.\ Rev.\ D {\bf 93}, no. 11, 113007 (2016)
  [arXiv:1601.02609 [hep-ph]].
   
\bibitem{Weinberg:2013kea} 
  S.~Weinberg,
  Phys.\ Rev.\ Lett.\  {\bf 110}, no. 24, 241301 (2013)
  doi:10.1103/PhysRevLett.110.241301
  [arXiv:1305.1971 [astro-ph.CO]].
   
\bibitem{Cheung:2017lpv} 
  K.~Cheung, H.~Okada and Y.~Orikasa,
  arXiv:1706.02084 [hep-ph].

   
\bibitem{Kanemura:2012rs} 
  S.~Kanemura and K.~Yagyu,
  Phys.\ Rev.\ D {\bf 85}, 115009 (2012)
  doi:10.1103/PhysRevD.85.115009
  [arXiv:1201.6287 [hep-ph]].


\bibitem{pdg} 
  C.~Patrignani {\it et al.} [Particle Data Group],
  Chin.\ Phys.\ C {\bf 40}, no. 10, 100001 (2016).

\bibitem{Kanemura:2012rj} 
  S.~Kanemura and H.~Sugiyama,
  Phys.\ Rev.\ D {\bf 86}, 073006 (2012)
  [arXiv:1202.5231 [hep-ph]].
  
\bibitem{Nomura:2016dnf} 
  T.~Nomura, H.~Okada and Y.~Orikasa,
  Phys.\ Rev.\ D {\bf 94}, no. 11, 115018 (2016)
  [arXiv:1610.04729 [hep-ph]].

\bibitem{Okada:2015nca} 
  H.~Okada and Y.~Orikasa,
  Phys.\ Rev.\ D {\bf 93}, no. 1, 013008 (2016)
  [arXiv:1509.04068 [hep-ph]].

  \bibitem{Dev:2012sg} 
  P.~S.~B.~Dev and A.~Pilaftsis,
  Phys.\ Rev.\ D {\bf 86}, 113001 (2012)
  [arXiv:1209.4051 [hep-ph]].  

  \bibitem{Dev:2012bd}
  P.~S.~Bhupal Dev and A.~Pilaftsis,
  Phys.\ Rev.\ D {\bf 87} (2013) no.5,  053007
  [arXiv:1212.3808 [hep-ph]].  

  
\bibitem{Das:2017ski} 
  A.~Das, T.~Nomura, H.~Okada and S.~Roy,
  Phys.\ Rev.\ D {\bf 96}, no. 7, 075001 (2017)
  [arXiv:1704.02078 [hep-ph]].



\bibitem{Okada:2012np} 
  H.~Okada and T.~Toma,
  Phys.\ Rev.\ D {\bf 86}, 033011 (2012)
  [arXiv:1207.0864 [hep-ph]].


\bibitem{Fernandez-Martinez:2016lgt} 
  E.~Fernandez-Martinez, J.~Hernandez-Garcia and J.~Lopez-Pavon,
  JHEP {\bf 1608}, 033 (2016)
  [arXiv:1605.08774 [hep-ph]].
  
\bibitem{Agostinho:2017wfs} 
  N.~R.~Agostinho, G.~C.~Branco, P.~M.~F.~Pereira, M.~N.~Rebelo and J.~I.~Silva-Marcos,
  arXiv:1711.06229 [hep-ph].
  
  
\bibitem{Kallosh:1995hi} 
  R.~Kallosh, A.~D.~Linde, D.~A.~Linde and L.~Susskind,
  Phys.\ Rev.\ D {\bf 52}, 912 (1995)
  [hep-th/9502069].
  
  
  
\bibitem{Kim:2008hd} 
  J.~E.~Kim and G.~Carosi,
  Rev.\ Mod.\ Phys.\  {\bf 82}, 557 (2010)
  [arXiv:0807.3125 [hep-ph]].
  
  
\bibitem{Ayala:2014pea} 
  A.~Ayala, I.~Dom\'inguez, M.~Giannotti, A.~Mirizzi and O.~Straniero,
  Phys.\ Rev.\ Lett.\  {\bf 113}, no. 19, 191302 (2014)
  [arXiv:1406.6053 [astro-ph.SR]].
  
\bibitem{Sedrakian:2015krq} 
  A.~Sedrakian,
  Phys.\ Rev.\ D {\bf 93}, no. 6, 065044 (2016)
  [arXiv:1512.07828 [astro-ph.HE]].

  
\bibitem{Arvanitaki:2014dfa} 
  A.~Arvanitaki and A.~A.~Geraci,
  Phys.\ Rev.\ Lett.\  {\bf 113}, no. 16, 161801 (2014)
  [arXiv:1403.1290 [hep-ph]].

  
\bibitem{Sushkov:2011zz} 
  A.~O.~Sushkov, W.~J.~Kim, D.~A.~R.~Dalvit and S.~K.~Lamoreaux,
  Phys.\ Rev.\ Lett.\  {\bf 107}, 171101 (2011)
  [arXiv:1108.2547 [quant-ph]].
  
  
\bibitem{Perez:2008ha} 
  P.~Fileviez Perez, T.~Han, G.~y.~Huang, T.~Li and K.~Wang,
  Phys.\ Rev.\ D {\bf 78}, 015018 (2008)
  [arXiv:0805.3536 [hep-ph]].

  
\bibitem{Chiang:2012dk} 
  C.~W.~Chiang, T.~Nomura and K.~Tsumura,
  Phys.\ Rev.\ D {\bf 85}, 095023 (2012)
  [arXiv:1202.2014 [hep-ph]].
  
\bibitem{Kanemura:2014ipa} 
  S.~Kanemura, M.~Kikuchi, H.~Yokoya and K.~Yagyu,
  PTEP {\bf 2015}, 051B02 (2015)
  [arXiv:1412.7603 [hep-ph]].

  
\bibitem{Ghosh:2017pxl} 
  D.~K.~Ghosh, N.~Ghosh, I.~Saha and A.~Shaw,
  Phys.\ Rev.\ D {\bf 97}, no. 11, 115022 (2018)
  doi:10.1103/PhysRevD.97.115022
  [arXiv:1711.06062 [hep-ph]].

  
\bibitem{Du:2018eaw} 
  Y.~Du, A.~Dunbrack, M.~J.~Ramsey-Musolf and J.~H.~Yu,
  JHEP {\bf 1901}, 101 (2019)
  [arXiv:1810.09450 [hep-ph]].

      \bibitem{Belyaev:2012qa} 
  A.~Belyaev, N.~D.~Christensen and A.~Pukhov,
  Comput.\ Phys.\ Commun.\  {\bf 184}, 1729 (2013)
  [arXiv:1207.6082 [hep-ph]].
  
\bibitem{Nadolsky:2008zw} 
  P.~M.~Nadolsky, H.~L.~Lai, Q.~H.~Cao, J.~Huston, J.~Pumplin, D.~Stump, W.~K.~Tung and C.-P.~Yuan,
  Phys.\ Rev.\ D {\bf 78}, 013004 (2008)
  [arXiv:0802.0007 [hep-ph]].
  
\bibitem{Aad:2014hja} 
  G.~Aad {\it et al.} [ATLAS Collaboration],
  JHEP {\bf 1508}, 138 (2015)
  [arXiv:1411.2921 [hep-ex]].
    
\bibitem{Alwall:2014hca} 
  J.~Alwall {\it et al.},
  JHEP {\bf 1407}, 079 (2014)
  [arXiv:1405.0301 [hep-ph]].

  \bibitem{Alloul:2013bka} 
  A.~Alloul, N.~D.~Christensen, C.~Degrande, C.~Duhr and B.~Fuks,
  Comput.\ Phys.\ Commun.\  {\bf 185}, 2250 (2014)
  [arXiv:1310.1921 [hep-ph]].
  
\bibitem{Deans:2013mha} 
  C.~S.~Deans [NNPDF Collaboration],
  arXiv:1304.2781 [hep-ph].
  
  \bibitem{Ref:Pythia}
  T.~Sjostrand, S.~Mrenna, P.~Z.~Skands,
  JHEP {\bf 0605 },  026 (2006).



\bibitem{Ref:PGS}
\url{http://conway.physics.ucdavis.edu/research/software/pgs/pgs4-general.htm}.

\end{thebibliography}
\end{document}